%% file: IEEE-conference-template-062824.tex
\documentclass[conference]{IEEEtran}
\IEEEoverridecommandlockouts

\usepackage{cite}
\usepackage{amsmath,amssymb,amsfonts}
\usepackage{algorithmic}
\usepackage{graphicx}
\usepackage{textcomp}
\usepackage{xcolor}
\usepackage{booktabs}
\usepackage{url}
\usepackage{hyperref}
\usepackage[normalem]{ulem}

\def\BibTeX{{\rm B\kern-.05em{\sc i\kern-.025em b}\kern-.08em
    T\kern-.1667em\lower.7ex\hbox{E}\kern-.125emX}}

\begin{document}

\title{Generation of Musical Timbres using a Text-Guided Diffusion Model

\thanks{$^{1}$ Chair of Computer Vision and Artificial Intelligence, TU Munich}

\thanks{$^{2}$ Munich Center for Machine Learning}

}

\author{\IEEEauthorblockN{Weixuan Yuan$^{1}$}
\and
\IEEEauthorblockN{Qadeer Khan$^{1,2}$}
\and
\IEEEauthorblockN{Vladimir Golkov$^{1,2}$}
}
\maketitle

\input{sections/Abstract}


\input{sections/Introduction}
\input{sections/Related_Work}

\input{sections/Method}

\input{sections/Experiments}

\input{sections/Conclusions}

\end{document}

%% file: sections/Abstract.tex
\begin{abstract}

In recent years, text-to-audio systems have achieved remarkable success, enabling the generation of complete audio segments directly from text descriptions. While these systems also facilitate music creation, the element of human creativity and deliberate expression is often limited.
In contrast, the present work allows composers, arrangers, and performers to create the basic building blocks for music creation: audio of individual musical notes for use in electronic instruments and DAWs. Through text prompts, the user can specify the timbre characteristics of the audio.
We introduce a system that combines a latent diffusion model and multi-modal contrastive learning to generate musical timbres conditioned on text descriptions. By jointly generating the magnitude and phase of the spectrogram, our method eliminates the need for subsequently running a phase retrieval algorithm, as related methods do.

Audio examples, source code, and a web app are available at \url{https://wxuanyuan.github.io/Musical-Note-Generation/}

\end{abstract}

%% file: sections/Introduction.tex
\section{Introduction}

Composers, arrangers, and performers require tools that generate the audio of musical notes, the fundamental building blocks of musics, of specific timbre. For example, they might need a finely controlled \emph{"a soft sound played by an orchestral instrument"}, as opposed to entire music segments. In recent years, leveraging diffusion models for text-guided audio generation has yielded remarkable results such as \textit{DiffSound, AudioLDM, AudioLM, MusicLM, MeLoDy} \cite{Diffsound,Audioldm,borsos2023audiolm,Audioldm2,agostinelli2023musiclm,lam2024efficient} etc. While these advancements provide convenient means for general users to access audio and even generate music, they are not explicitly designed to assist musicians in shaping timbre, which is essential for crafting melodies and harmonies. Instead, these methods often create the impression that musicians are becoming redundant.

Therefore, to support, rather than replace musicians, we propose a framework specifically targeted towards timbre generation. This simplifies the complex, experience-dependent, and potentially costly process of obtaining audio of musical notes with desired timbres. This typically involves selecting, purchasing, and adjusting synthesizers which can be avoided with our method; thereby providing significant convenience for musical arrangement.

In contrast to other methods for audio/music generation, our diffusion-based model is the first to overcome dataset limitations and achieve smooth timbre control guided by natural language at the musical note level, with state-of-the-art performance. Furthermore, it empowers human arrangers with flexibility in timbre manipulation, fostering a more creative music composition process and enabling the generation of previously non-existent timbres.

In summary, our contributions include the following:
\begin{enumerate}
\itemsep0em 
\item Allows composers, arrangers, and performers fine-grained control in timbre creation at the musical Note level. 
\item Enables joint generation of the magnitude and phase of the spectrogram, eliminating the need for subsequently running a phase retrieval algorithm.
\item Enables parameterized timbre modification guided by natural language, validated through quantitative experiments,
\item We shall provide the entire workflow and all models for assisting musicians upon acceptance of the paper.
\end{enumerate}

The sample audio and source code can be found online at \url{https://wxuanyuan.github.io/Musical-Note-Generation/}

\begin{figure}[!ht]
    \centering
    \includegraphics[page=14, width=1.\linewidth, trim={15mm, 55mm, 70mm, 22mm}, clip]{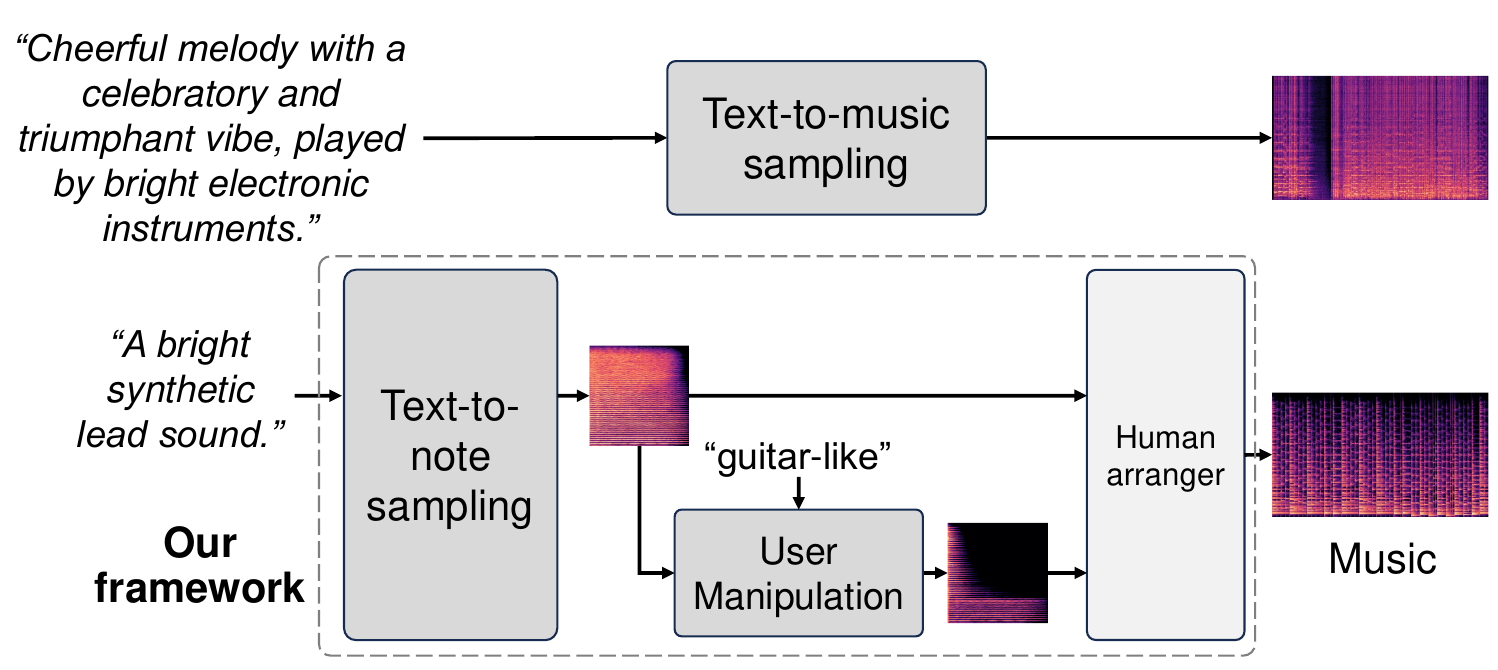}
    \caption{Difference between text-to-sound systems  (Top) and the proposed method (Bottom). Unlike end-to-end text-to-sound systems, human creativity in musical arrangement is preserved. Our method first generates fixed-length musical notes with the desired timbre based on text descriptions (\ref{sec:note_generation}). This generated note can optionally be modified by the user for more fine-grained control of desired output. After that, the fixed-length notes are adjusted to varying lengths using the diffusion-based inpainting method \cite{RePaint}. All three stages use the same models, trained solely on fixed-length samples. Finally, these different notes are arranged by human musicians to create music.
    }
    \label{fig: comparison}
\end{figure}

%% file: sections/Related_Work.tex
\begin{figure*}[!ht]
    \centering
    \includegraphics[page=2, width=0.7\linewidth, trim={2mm, 4mm, 4.5mm, 2mm}, clip]{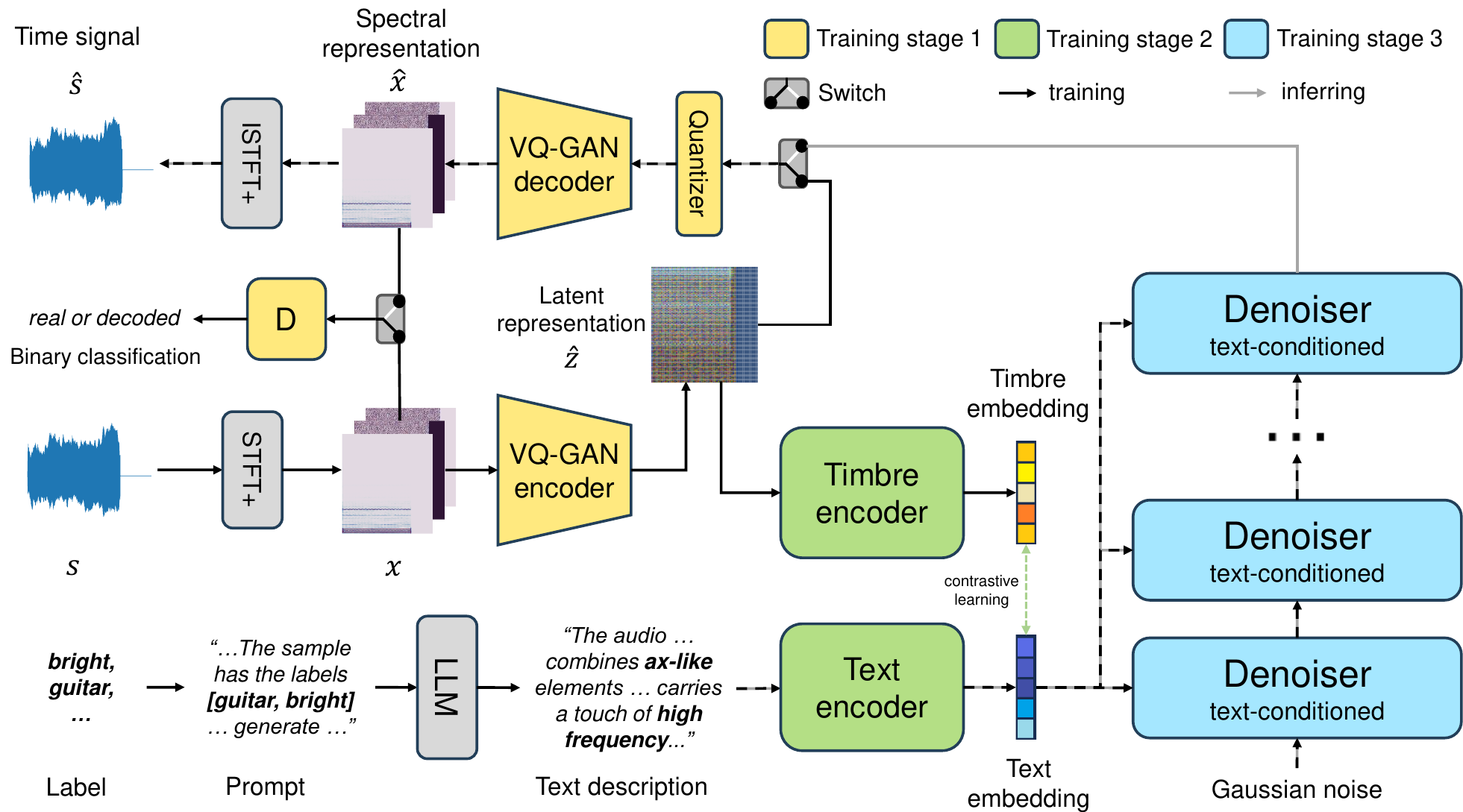}
    \caption{
    Architecture overview of our framework for generating timbre. It combines multi-modal contrastive learning and latent diffusion models. STFT+ and ISTFT+ represent the non-trainable time-frequency domain transformations of audio signals~\(S\).
    A pretrained LLM is used to augment labels such as ``bright, guitar'' from the NSynth dataset to diverse text descriptions. The training is divided into three phases:
    (1)~A~VQ-GAN (in yellow) is trained as an autoencoder for the spectral representation of real samples.
    Its discriminator~\(D\) is trained to distinguish spectral representations of real samples (i.e.~\(x\) for all training samples) from those of generated samples (i.e.~\(\hat{x}\) for all training samples). The encoder, decoder, and quantizer are trained to fool the discriminator, i.e.~to produce realistic~\(\hat{x}\).
    (2)~A~text encoder (pretrained using CLAP \cite{CLAP}) and a timbre encoder (both shown in green) are trained to map text descriptions and the timbre representation \(\hat{z}\) into a unified embedding space via contrastive learning.
    (3)~A diffusion model (in blue) is trained to produce latent representations conditioned by the text embeddings. During the inference stage, the output of the diffusion model is passed to the VQ-GAN decoder. Details of the individual components are provided in \ref{sec:note_generation}. For further details on the model components, including hyperparameters and training settings, please refer to the \href{https://wxuanyuan.github.io/Musical-Note-Generation/}{project page}.
    }
    \vspace{-1em}
    \label{fig:architecture}
\end{figure*}

\section{Related Work}

\noindent{\textbf{Diffusion Models:}}
Diffusion models have been demonstrated to be effective in image generation \cite{DDPM}. In particular, Latent Diffusion Models (LDM) \cite{StableDiffusion} improved further by operating in the latent space, significantly reducing computational costs while maintaining high-quality outputs. Moreover, diffusion-based zero-shot methods have emerged, such as image style transfer and inpainting \cite{RePaint}. These methods rely solely on pretrained models without the need for additional training.

\noindent{\textbf{Text-to-audio Systems}}
Despite the recent lack of advancements in timbre generation on the musical note level, significant progress has been made in the broader field of audio generation. DiffSound \cite{Diffsound} utilizes diffusion models to generate audio, including music, sounds, and speech, conditioned by text embeddings obtained from a pretrained CLIP text encoder. AudioLDM \cite{Audioldm} enhances generation quality by employing contrastive learning to fine-tune the CLAP text encoder \cite{CLAP}. Additionally, AudioLDM explores advanced techniques such as text-guided audio-to-audio style transfer and inpainting in the spectral space. Furthermore, AudioLM leveraged language modeling methods to achieve long-term and consistent audio generation \cite{borsos2023audiolm}. Attempts have been made for music specifically, such as MusicLM and MeLoDy \cite{agostinelli2023musiclm,lam2024efficient}. These methods operate on the audio or music-segment level, while we focus on the note level.

%% file: sections/Method.tex
\section{Method}

This section describes the details of text-conditioned
Timbre generation (Subsection \ref{sec:note_generation}, Figure \ref{fig:architecture}) and Timbre Manipulation (Subsection \ref{sec:note_manipulation_methods}).

\subsection{Text-conditioned Timbre Generation}
\label{sec:note_generation}
\noindent{\textbf{Latent Representation on the Audio of Musical Notes:}}

This work aims to generate the spectrogram of a musical note with a timbre specified through a text prompt. To expedite training, we train the diffusion model on the lower-dimensional latent representation of the audio.
The audio is first converted to a spectral representation \(x \in \mathbb{R}^{3 \times H \times W}\), where the three channels correspond to log-magnitude, sine phase, and cosine phase, respectively.
We do not use mel-scaled spectrograms due to their tendency to compress high-frequency information \cite{natsiou2021audio}, which is detrimental to high-quality music synthesis.

This spectral representation is compressed to and reconstructed from a latent representation \(\hat{z} \in \mathbb{R}^{C \times \frac{H}{r} \times \frac{W}{r}}\), where \(C\) represents the number of channels, and \(r\) denotes the spatial compression scale, via a VQ-GAN \cite{VQGAN}. As depicted in the yellow section of Figure \ref{fig:architecture},

\noindent{\textbf{Textual Description Augmentation: }}
\label{sec:data_augmentation}
To facilitate the creation of virtual instruments guided by textual descriptions, timbre-text pairs are essential. NSynth is a high-quality sound dataset with high timbre diversity, where each sound sample is annotated to an instrument ID, an instrument source, an instrument type, and various timbre qualities \cite{NSynth}. However, these annotations are not in legible English. Therefore, to accommodate a wide range of potential text descriptions, GPT3.5 \cite{GPT3} API offered by \cite{openai_gpt3.5_turbo} is employed to transform these labels into different types of text descriptions, including concatenation of keywords, natural language description, and short phrases.

\noindent{\textbf{Contrastive Representation Learning: }}
\label{sec:contrastive_representationlearning}
The multi-modal nature of our approach necessitates a shared representation between text and timbre. To ensure this, we train a timbre encoder and a text encoder, which respectively map the latent audio representation and text descriptions to their corresponding embeddings within a unified latent space.
This is achieved using contrastive loss introduced in \cite{ContrastiveLossRef1, CLIP}.

\noindent{\textbf{Denoising Diffusion on Latent Representations: }}
The generation of latent representations \(\hat{z} \) is modeled by the reverse process of DDPM \cite{DDPM, StableDiffusion}:
\begin{align}
p_{\theta}(z_{0:T}|e^t) := p(z_T) \prod_{t=1}^{T} p_{\theta}(z_{t-1}|z_t,e^t),
\end{align}
where \(p(z_T) := \mathcal{N}(0, I)\) and \(e^t\) is the text embedding. The intermediate transitions are parameterized by estimations of a neural network:
\begin{align}
p_{\theta}(z_{t-1}|z_t,e^t) := \mathcal{N}(z_{t-1}; \mu_{\theta}(z_t, t, e^t), \Sigma_{\theta}(z_t, t, e^t)).
\end{align}
Model parameters \(\theta\) are optimized by minimizing an adapted version of the variational lower bound of the negative log-likelihood \cite{VariationalLowerBound}:
\begin{equation}
L_{\textnormal{t}}^{\textnormal{simple}} = \mathbb{E}_{t, x_0, \boldsymbol{\epsilon}} \left[ \left\| \epsilon - \boldsymbol{\epsilon}_{\theta}(z_t, t, e^t) \right\|^2 \right].
\end{equation}
Moreover, classifier-free guidance \cite{CFG} is applied during training, which involves randomly replacing the text embedding \(e^t\) with the embedding of an empty string with probability \( p \).

\subsection{Timbre Manipulation Methods}
\label{sec:note_manipulation_methods}

Due to the parallels between spectral representations and images, existing diffusion-model-based techniques in computer vision can facilitate the transformation and manipulation of musical timbres. For instance, the RePaint \cite{RePaint} method generates diverse content within designated regions while maintaining the overall naturalness of the image. This is achieved by replacing areas in the denoised representation at each time step, guided by a fixed mask, with the noise-added representation of the original image.

RePaint can be adapted for audio and perform localized modifications to an input sound within the spectral domain \cite{Audioldm}. Using a mask that covers the entire representation enables global modifications, transforming the input sound in alignment with the text description. As illustrated in Figure \ref{fig: comparison}, such timbre manipulation methods are inserted into our framework as an optional stage before human arrangement. This stage accepts any input sound, including outputs from text-to-audio systems.

%% file: sections/Experiments.tex
\section{Experiments}

Experiments were conducted on the NSynth dataset \cite{NSynth}, the largest open-source dataset of musical note audio samples. For comparison, we additionally compare against the following models: a)~\emph{\textbf{GAN}}, a generative adversarial network for audio synthesis inspired by \cite{GANSynth}, but using the noise predictor architecture of our method as the generator; b)   ~\emph{\textbf{AudioLDM}}: a text-to-sound-system that was trained on more general audio datasets. c)~\emph{\textbf{AudioLDM\_A}}: same as AudioLDM\cite{Audioldm} but adapted to our musical note audio dataset; d)~\emph{\textbf{Ours\_C}}: same as our framework but using pretrained CLAP as text-encoder.

It is pertinent to highlight that like our method, AudioLDM is also a text-conditioned diffusion model that provides the code, which we used to force the model to produce audio of musical notes rather than audio using appropriate text prompts. This allows our method to be compared with AudioLDM. However, this may not lead to a fair evaluation since AudioLDM was originally trained to generate general audio rather than musical notes. This is why, we  adapted the  AudioLDM model by additionally training on the NSynth dataset.

\noindent{\textbf{Quantitative Evaluation: }} Quantitative results of the unconditioned and text-conditioned sampling of the different generative models are compared in \ref{tab:uncondintioned_sampling} \ref{tab:conditioned_sampling_results} respectively.
Based on the results of objective metrics including Fréchet Audio Distance (\emph{FAD}), Precision-and-Recall, and Inception Score (\emph{IS}), it is evident that our method demonstrates superior performance in terms of realism and diversity. The pretrained Audio Transformer \cite{AST} was employed for feature extraction. These results also validate the importance of individual components in our framework, as demonstrated by the significant performance drop observed when key elements such as the diffusion model, its architecture, or the contrastive learning approach are removed. Please note that the unusually high FAD can be attributed to the inherent characteristics of the dataset used, rather than any flaws in the experimental setup.

\begin{table}[!ht]
\centering
\small
\setlength{\tabcolsep}{4pt}
\begin{tabular}{lcccccc}
\toprule
\textbf{Model} & \textbf{IS} $\uparrow$ & \textbf{FAD} $\downarrow$ & \textbf{Prec.} $\uparrow$ & \textbf{Recall} $\uparrow$ & \textbf{Size}                 \\ \midrule
GAN & 30.1 & 636.0 & 0.64 & 0.00 & 107M  \\ \midrule
AudioLDM & 6.1 & 1127.6 & 0.02 & 0.20 & 857M \\ \midrule
AudioLDM\_A & 28.5 & 420.5 & 0.70 & 0.20 & 857M \\ \midrule
Ours\_C & 206.1 & 316.5 & 0.79 & 0.36 & 107M \\ \midrule
Ours & \textbf{266.1} & \textbf{268.0} & \textbf{0.81} & \textbf{0.50} & 107M   \\ \bottomrule
\end{tabular}
\caption{Comparison of unconditioned sampling results across different models.}
\label{tab:uncondintioned_sampling}
\end{table}

\begin{table}[!ht]
\centering
\small
\setlength{\tabcolsep}{6pt}
\begin{tabular}{lcccc}
\toprule
\textbf{Model} & \textbf{IS} $\uparrow$ & \textbf{FAD} $\downarrow$ & \textbf{Prec.} $\uparrow$ & \textbf{Recall} $\uparrow$ \\ \midrule
GAN & 25.5 & 763.2 & 0.25 & 0.01 \\ \midrule
AudioLDM & 3.7 & 982.2 & 0.10 & 0.01 \\ \midrule
AudioLDM\_A & 19.4 & 516.6 & 0.30 & 0.44 \\ \midrule
Ours\_C & \textbf{92.7} & 444.5 & 0.41 & 0.62 \\ \midrule
Ours & 86.8 & \textbf{302.8} & \textbf{0.55} & \textbf{0.66} \\ \bottomrule
\end{tabular}
\caption{Comparison of text-conditioned sampling results across different models.}
\label{tab:conditioned_sampling_results}
\end{table}

\noindent{\textbf{Qualitative Human Evaluation: }}
A qualitative human evaluation has been conducted to support the conclusions drawn from the quantitative objective evaluation. All 19 evaluators had a music background, with 7 having professional music training.
Each participant was asked to hear a number of paired audio sequences and decide which of the two in the pair was better in terms of quality, text synchronization, and timbre consistency throughout the audio.  One of the audio sequences in the pair was generated with our method and the other used the AudioLDM and its adaptive version. The results are shown in Table \ref{tab:subjective}. It can be seen from the first two rows that our method outperforms both baselines, thereby corroborating the conclusion of the quantitative evaluation.

\begin{table}[!ht]
\centering
\small
\setlength{\tabcolsep}{2pt}
\caption{Qualitative evaluation results measuring sound quality and text synchronization. A large proportion of participants find our method to be better on these metrics when compared with both \emph{AudioLDM} and \emph{AudioLDM\_A}. }
\label{table:human_evaluation}
\begin{tabular}{llr}
\toprule
                           & quality / text sync.  \\                  
\midrule
Ours vs. AudioLDM    & 73\% \,\,\,\, / \,\,\,\,\, 90\% \,\,\,\,\, \\ \hline
Ours vs. AudioLDM\_A & 73\% \,\,\,\, / \,\,\,\,\, 80\% \,\,\,\,\,  \\ \hline
\end{tabular}
\label{tab:subjective}
\end{table}

\begin{figure}[]
    \centering
    \includegraphics[page=3, width=1\linewidth, trim={30mm, 86mm, 60mm, 47mm}, clip]{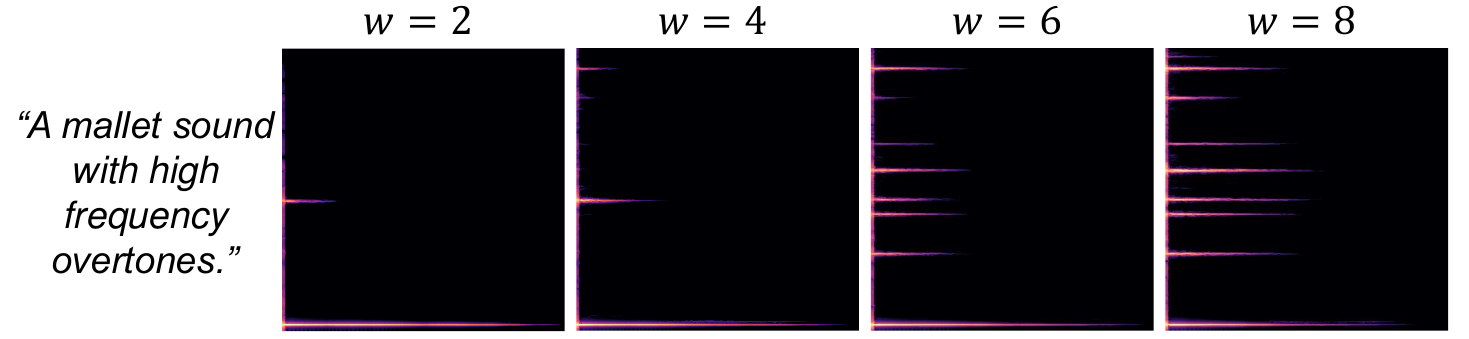}
    \caption{Results of conditioned sampling with varying guidance scales \(w\). As \(w\) increases, more high-frequency components are introduced into the spectrogram in line with the text description.
    }
    \label{fig:CFG_examples}
\end{figure}

\begin{figure}[!ht]
    \centering
    \includegraphics[page=5, width=1\linewidth, trim={47mm, 60mm, 60mm, 50mm}, clip]{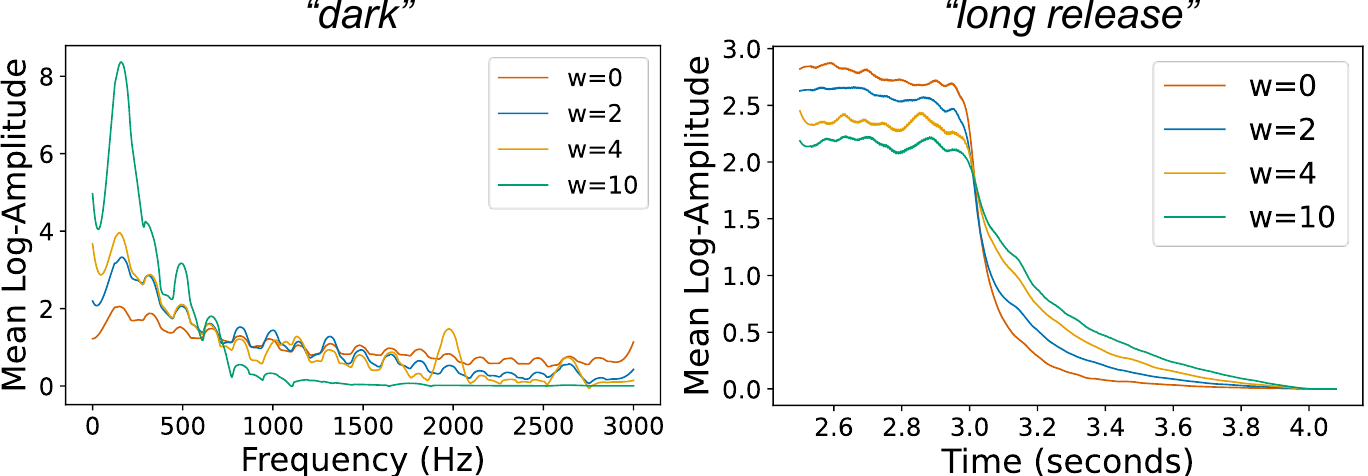}
    \caption{Variations of the mean amplitude distribution along frequency (left) and time (right) dimensions at different guidance scales \(w\). With the increase of guidance scale \(w\), the amplitude distribution varies as required by the text description; specifically, ``dark'' for the lower frequency range, and ``long release'' for the release stage of the sound.}
    \label{fig:mean_amplitude_vs_cfg}
\end{figure}

\noindent{\textbf{Effect of Guidance Scale: }}
The influence of text on the final generated output is controlled by the guidance factor $w$. This is illustrated in  \ref{fig:CFG_examples} where for greater $w$ values, we see higher frequency components appearing in the spectrogram to reflect the text description better. 
Meanwhile, the dynamic effects of the guidance scale are visualized in the mean amplitude graphs in \ref{fig:mean_amplitude_vs_cfg}.
The left panel shows that as the guidance scale for the ``dark'' text descriptions increases, the amplitude distribution progressively shifts toward the lower frequency range. Similarly, in the right panel, for the ``long release'' text description, the amplitude of the release stage (the final second) increases with an increasing guidance scale.

\begin{figure}[!ht]
    \centering
    \includegraphics[page=14, width=1\linewidth, trim={29mm, 118mm, 104mm, 25mm}, clip]{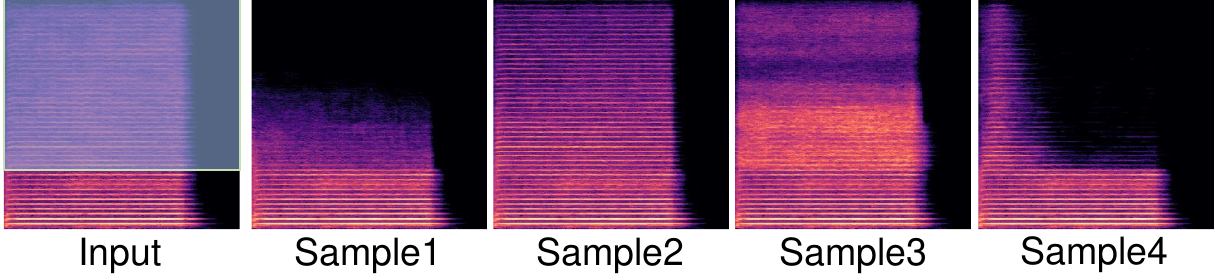}
    \caption{Timbre inpainting examples. The modified regions are highlighted with masks bordered in light blue. The text description is an empty string.}
    \label{fig:repaint_results}
\end{figure}

\begin{figure}[!ht]
    \centering
    \includegraphics[page=9, width=1\linewidth, trim={65mm, 50mm, 71mm, 60mm}, clip]{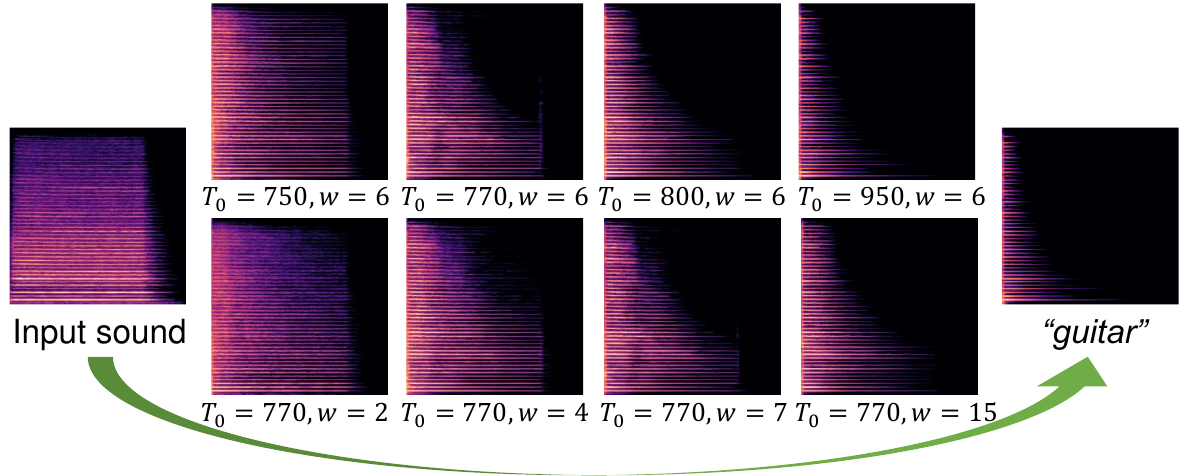}
    \caption{Timbre transformation examples. Smooth transitions in timbre are achieved by altering the guidance scale \(w\) (upper row), or by changing the noising strength through the initial time step \(T_0\) (lower row). The text description is ``guitar''.}
    \label{fig:timbre_transfer}
\end{figure}

\noindent{\textbf{Text-guided Timbre Modification: }} Figure \ref{fig:repaint_results} showcases examples of localized spectrogram modifications using in-painting, where the high-frequency regions are diversely re-generated while maintaining the naturalness with the unchanged low-frequency parts. Figure \ref{fig:timbre_transfer} presents the results of global spectrogram transformations. The transformation outcomes can  smoothly be controlled either by fine-tuning the guidance scale or noising strength thereby offering practical significance for arrangers for achieving the desired timbre. Notably, these methods enable the creation of unique and even non-existent timbres. We recommend visiting the  \href{https://wxuanyuan.github.io/Musical-Note-Generation/}{project page} to listen to the results or directly explore our \href{https://huggingface.co/spaces/WeixuanYuan/DiffuSynthV0.2}{web app} for a hands-on experience.

%% file: sections/Conclusions.tex
\section{Conclusions}
\label{section: Conclusions}

We introduced a novel method of generating and manipulating musical timbres via text prompts. This allows advancing the usability and flexibility of deep generative models in arrangement by musicians. The generation/manipulation of musical timbres is faithful to the text prompts, the intensity of which can be smoothly controlled with parameters such as the guidance scale.